\documentclass{elsart}
\usepackage{psfig}

\begin{document}

\begin{frontmatter}
  \title{ $\omega$- and $\phi$-meson production in $p n \to d M$
    reactions near threshold and OZI-rule violation}

\author[INR]{V.Yu. Grishina\thanksref{DFG}},
\author[ITEP]{L.A. Kondratyuk\thanksref{DFG}}  and
\author[Juelich]{M. B\"uscher\thanksref{DFG}}

\address[INR]{Institute for Nuclear Research, 60th October Anniversary
  Prospect 7A, 117312 Moscow, Russia}
\address[ITEP]{Institute of Theoretical and Experimental Physics, B.\
  Cheremushkinskaya 25, 117259 Moscow, Russia}
\address[Juelich]{Forschungszentrum J\"ulich, Institut f\"ur
  Kernphysik, 52425 J\"ulich, Germany}
\thanks[DFG]{Supported by DFG and RFFI.}

\begin{abstract}
  We investigate the reactions $p n \to d \omega$ and $p n\to d \phi$
  close to the corresponding thresholds. The $S$-wave amplitudes are
  calculated within the framework of the two-step model which is
  described by a triangle graph with $\pi$, $\rho$ and $\omega$ mesons
  in the intermediate state.  The cross sections of the reactions $p n
  \to d \omega$ and $p n\to d \phi$ are predicted to be significantly
  larger than the cross sections of the corresponding reactions $p p
  \to pp \omega$ and $p p\to pp\phi$ at the same values of the c.m.\
  excess energy $Q$. The ratio of the yields of $\phi$ to $\omega$
  is found to be $(30 \pm 7) \times 10^{-3}$.
  

\end{abstract}
\end{frontmatter}
\section{Introduction}
It is well known (see e.g.\ \cite{Lipkin,Ellis1,Ellis2}) that the
ratio of the $\phi /\omega$ yields 
\begin{equation}
  R = \frac{\sigma_{A+B\to \phi X}}{\sigma_{A+B\to \omega X}}\ ,
\label{eq:phitoomega}
\end{equation}
is a particularly sensitive probe of the OZI rule \cite{Okubo}. Using
the standard value for the deviation $\delta = \theta -
\theta_{\mathrm{i}} = 3.7^{\circ}$ from the ideal SU(3)$_f$ mixing
angle $\theta_{\mathrm{i}} = 35.3^{\circ}$ we have $R/f = 4.2\times
10^{-3}$ \cite{Ellis2}, where $f$ is the ratio of the phase-space
factors.  However, experimental data show an apparent excess of $R/f$
above the standard value which varies from $(10-30)\times 10^{-3}$ in
$\pi N$ and $NN$ collisions to $(100-250)\times 10^{-3}$ in $\bar{N}N$
annihilation at rest and in flight (see e.g.\ the discussion in
\cite{Ellis2}).  In Ref.\cite{Ellis2} the large excess of $R$ in $pp$
and $\bar{p}p$ collisions over the prediction by the OZI rule was
treated in terms of ``shake-out'' and ``rearrangement'' of an
intrinsic $\bar{s}s$ component in the nucleon wave function. On the
other hand, in papers \cite{Locher,Buzatu} the strong violation of the
OZI rule in $\bar{p}p$ annihilation at rest was explained in terms of
hadronic intermediate $K\bar{K}^{\ast}$ states which might create
$\phi$ mesons.

Another argument in favor of a large admixture of hidden strangeness
in nucleons was related to an apparently large contribution of the
$\phi$-meson into the isoscalar spectral function which through the
dispersion relation defines the isoscalar nucleon form factor (see
Ref.\cite{Jaffe}). However, as it was shown later (see \cite{Meisner}
and references therein), the main contribution to the isoscalar
spectral function near 1 GeV stems from correlated $\pi \rho$ exchange
which does not involve strange quarks.

Therefore, the question whether there is a large admixture of hidden
strange\-ness in nucleons seems to be unclarified. Thus, it is
important to investigate such reactions where uncertainties in the
interpretation of $\omega$ and $\phi$ production in terms of
intermediate hadronic states are comparably small. In this paper we
argue that a good choice in this respect is the reaction
\begin{equation}
pn\to dM\ .
\label{eq:Vd}
\end{equation}
Here and below $M$ denotes the vector mesons $\omega$ and $\phi$.

We analyze contributions of hadronic intermediate states into the
$S$-wave amplitudes of the reactions $p n \to d \phi$ and $p n \to
d\omega$ within the framework of the two-step model (TSM) described by
triangle graphs with $\pi$-, $\rho$- and $\omega$-meson exchanges.
Previously this model was applied to the description of the Pontecorvo
reactions $\bar{p} d \to p M$ (see, e.g.,
Ref.\cite{kondrat1,kondrat2}). In a recent paper (see
Ref.\cite{Grishina}) it was demonstrated that the TSM can also
describe the cross section of the reaction $p n \to d \eta$ near
threshold with a reasonable choice of the coupling constants and
cut-off parameters for $\pi$-, $\rho$- and $\omega$-meson exchanges.
To predict the cross sections of the reactions $p n \to d\omega$ and
$p n \to d \phi$  we use a similar approach and the same set of
parameters for the $MNN$ coupling constants and cut-off parameters.
Note that if the $\phi$ and $\omega$ yields will be measured in
reaction (\ref{eq:Vd}) near threshold (which e.g.\ can be done at
COSY-J\"ulich), the results can be useful for a better understanding
of the OZI-rule violation dynamics.  For example, any significant
deviation from the prediction of the two-step model could be an 
evidence for the above mentioned ``shake out'' or ``rearrangement'' of
an intrinsic $\bar{s}s$ component in the nucleon wave function.

Note that recent measurements of the $\phi /\omega$ ratio in the
reaction $pd\to ^3\!\! HeX$ (performed at SATURNE II
\cite{Wurzinger1,Wurzinger2}) yield
\begin{equation}
  R/f = \left(63 \pm 5\ ^{+27}_{-8}   \right)\times 10^{-3}
  \label{eq:wurzinger}
\end{equation}
which is also clearly above the expectation 4.2$\times 10^{-3}$.
However the dynamics of the reaction $pd\to ^3\!\! HeX$ is yet not
well understood. According to \cite{Wilkin1} the two-step model
underestimates the SATURNE data by a factor 2, while according to
\cite{Uzikov} the discrepancy of the two-step model with the data
might be even larger when spin-effects are taken into account.

Experiments on $\omega$ and $\phi$ production in the reaction $ p p
\to p p M$ close to threshold were performed by the SPES3 and DISTO
collaborations at SATURNE \cite{SPES3,DISTO} (see also the
calculations of $\omega$ production in \cite{Speth}).  According to
the DISTO data the ratio of the $\phi/\omega$ production cross
sections at 2.85 GeV is $\sigma_{\mathrm{tot}}(p p \to p p \phi)$/
$\sigma_{\mathrm{tot}} (p p \to pp\omega) = (3.7 \pm 1.3)\times
10^{-3}$. Introducing corrections for phase-space effects the authors
of Ref.\cite{SPES3} found that in this case the $\phi/\omega$ ratio is
$(49 \pm 26)\times 10^{-3}$. Note that near threshold the
dynamics of the reactions $ p p \to p p M$, $ p n \to p n M$ and $ p n
\to d M$ are different because the first one is constrained by the
Pauli principle and the two protons in the final state should be in a
$^1 S_0$ state.  In the third case the final $pn$ system is in the $^3
S_1$ state while in the second case it can be in both states.
Therefore, a possible violation of the OZI rule is expected to be
different in all those cases.

Finally, another interesting point is that within the framework of the
line-reverse invariance (LRI) assumption the reaction $pn \to dM$ can
be related to the Pontecorvo reaction $\bar{p}d \to MN$.  The data
from the OBELIX and Crystal-Barrel collaborations result in a $\phi /
\omega $ ratio of about $(230 \pm 60)\times 10^{-3}$
\cite{Sapozhnikov,Crystal}. Therefore, if LRI is applicable we expect
the violation of the OZI rule in the reaction $pn \to dM$ to be much
larger than it is predicted by the two-step model, which assumes the
dominance of the hadronic intermediate states.

The paper is organized as follows. In Sect.~\ref{sec:twostep} we
derive the amplitudes of the reactions $pn\to d\phi$ and $pn\to
d\omega$ near threshold within the framework of the two-step model. In
Sect.~\ref{sec:results} we discuss the choice of parameters and
present the results of calculations.  Sect.~\ref{sec:conc} contains
our conclusions.

\section{The non-relativistic two-step model for the reaction  $pn\to d
  M$}
\label{sec:twostep}

\begin{figure}[htb] 
  \begin{center}
    \leavevmode
    \psfig{file=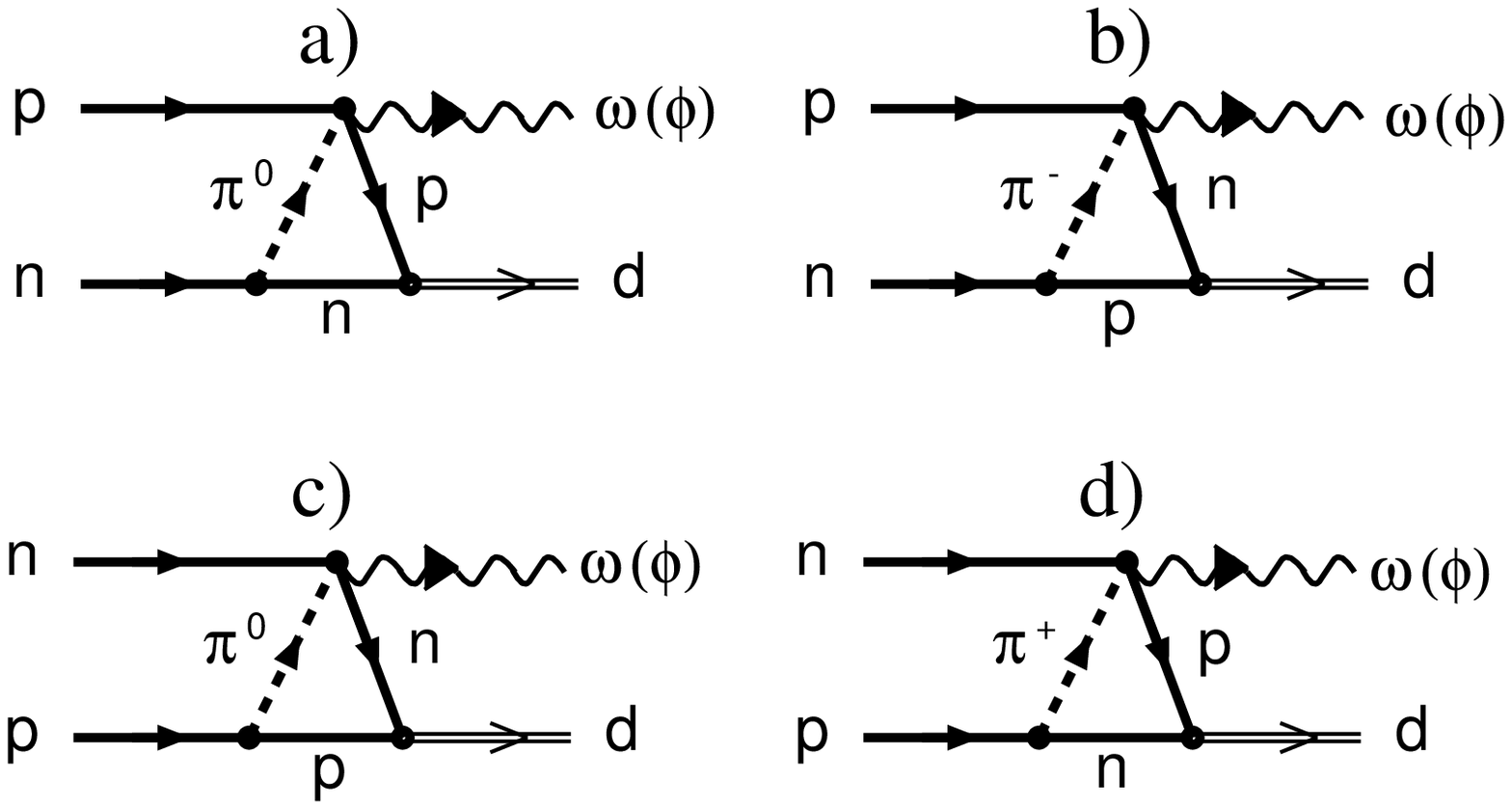,width=8.cm}
    \caption{Diagrams describing the two-step model (TSM). Note that
      besides the $\pi$-exchange contribution also diagrams involving
      the exchange of $\rho$ and $\omega$ mesons are taken into
      account.}
    \label{fig:tsm} 
  \end{center}
\end{figure}

The triangle diagrams describing the TSM are shown in
Fig.~\ref{fig:tsm}.  Besides the $\pi$ exchange we also take into
account $\rho$ and $\omega$ exchanges.

In the beginning let us consider the $\pi^0$-exchange term.  In order
to preserve the correct structure of the amplitude under permutations
of the initial nucleons (which should be symmetric in the isoscalar
state) the amplitude is written as the sum of the $t$- and $u$-channel
contributions in the following form
\begin{equation}
  T_{pn \to dM}^{\pi}(s,t,u)= A_{pn \to dM}^{\pi}(s,t)+ A_{pn \to
    dM}^{\pi}(s,u) \ ,
\label{Atu}
\end{equation}
where $M$ is the vector meson $\omega$ or $\phi$.  $s=(p_1+p_2)^2$,
$t=(p_3-p_1)^2$, and $u=(p_3-p_2)^2$ where $p_1$, $p_2$, $p_3$, and
$p_4$ are the 4-momenta of the proton, neutron, meson $M$ and
deuteron, respectively.  Since we are interested in the calculation of
the cross section of reaction (\ref{eq:Vd}) near  threshold where the
momenta of the deuteron and the meson are comparatively small, we can use
a non-relativistic description of those particles by neglecting the 4th
components of their polarization vectors.  The relative motion of
nucleons inside the deuteron is also treated non-relativistically.
Then one can write the two terms on the right hand side of
Eq.~(\ref{Atu}) as follows (see also \cite{Grishina})
\begin{eqnarray}
  A_{pn\to dM}^{\pi} (s,t) &=& \frac{f_{\pi}}{m_{\pi}}
  \varphi^{T}_{\lambda_{2}} ({\vec{p}}_2)\ (-i \sigma_2)
  {\vec{\sigma}}\cdot{\vec{M}}^{\pi}({\vec p}_1)\ {\vec{\sigma}}\cdot
  {\vec{\epsilon}} ^{\ast}_{d}\ {\vec{\sigma}}\cdot
  {\vec{\epsilon}} ^{\ast}_M\ \varphi_{\lambda_1} ({\vec{p}}_1)
  \ \times \nonumber \\
  &&  A_{\pi^0 N \to M N}(s_1,t) \ ,
\label{AVdt}
\end{eqnarray}

\begin{eqnarray}
  A_{pn\to d M}^{\pi} (s,u) &=& \frac{f_{\pi}}{m_{\pi}}
  \varphi^{T}_{\lambda_{1}} ({\vec{p}}_1)\ (-i \sigma_2)
  {\vec{\sigma}}\cdot{\vec{M}}^{\pi}(- {\vec p}_1 )\ {\vec{\sigma}}\cdot
  {\vec{\epsilon}} ^{\ast}_{d}\ {\vec{\sigma}}\cdot
  {\vec{\epsilon}} ^{\ast}_{M}\ \varphi_{\lambda_{2}} ({\vec{p}}_2)
  \ \times \nonumber \\
  &&  A_{\pi^0 N \to M N}(s_1,u) \ ,
\label{AVdu}
\end{eqnarray}

where $\vec{\epsilon}_d$ and $\vec{\epsilon}_M$ are the polarization
vectors of the deuteron and the meson; $\varphi_{\lambda}$ are the
spinors of the nucleons in the initial state, $m_{\pi}$ and $f_{\pi}$
are the pion mass and $\pi NN$ coupling constant.  The vector function
$\vec{M}^{\pi}({\vec p}_1 )$ is defined by the integral
\begin{eqnarray}
  &\vec{M}^{\pi}({\vec p}_1 ) &= \sqrt{2\,m} \int ({\vec{k}}+{\vec
    p}_1)\, \Phi_{\pi}({\vec{k}},{\vec{p}_1})\Psi_d(\vec{k})\,
  \frac{\mathrm{d}^3k}{(2\pi)^{3/2}} \ , \label{M} \\
  &\Phi_{\pi}({\vec{k}},{\vec{p_1}}) &=
  \frac{F_{\pi}(q^2)}{q^2-m_{\pi}^2} \ ,
 \label{ff}  
\end{eqnarray}
which contains the deuteron wave function  $\Psi_d(\vec{k})$ and the form
factor
at the $\pi NN$ vertex  $F_{\pi}(q^2)$. Other kinematical quantities which
are also 
dependent on  the momenta ${\vec p_1}$ and ${\vec k}$ are defined as
follows
\begin{eqnarray}
  &q^2 &= m_{\pi}^2 - \delta_0(\vec{k}^2+
  \beta(\vec{p}_1)) - 2{\vec {p}}_1 {\vec k} , \
  \vec{q} =  \vec{k}+\vec{p}_1 \ , \nonumber \\
  &\beta(\vec{p}_1) &=
  ({\vec{p}_1}^{\,2}+m_{\pi}^2-T_1^2)/\delta_0\ , \
  \delta_0 = 1+T_1/m, \ T_1= \sqrt{{\vec{p}_1}^2 + m^2} - m \ .
  \nonumber
\end{eqnarray}
with $m$ being the nucleon mass. 

Near threshold we take into account only the $S$-wave part of the
amplitude of the elementary reaction $\pi N \to M N$. Deriving
Eqs.(\ref{AVdt},\ref{AVdu}) we use the following spin structure of
the $\pi^0 N \to M N$ amplitude
\begin{eqnarray}
  \lefteqn{ \langle p_3^{\prime}\,\lambda^{\prime}_{3}; p_4^{\prime}
    \lambda_4^{\prime} | \hat{T}_{\pi N\to NM} |
    p_1^{\prime};p_2^{\prime} \lambda_2^{\prime}\rangle =}  \nonumber \\
  &&\varphi_{\lambda^{\prime}_{4}}^{\ast}(\vec{p}_4^{\,\prime})\ 
  \vec{\epsilon}_{\lambda^{\prime}_{3}}^{\ast \,(M)} \cdot
  \vec{\sigma}\ 
  \varphi_{\lambda^{\prime}_{2}}^{\ast}(\vec{p}_2^{\,\prime})\ A_{\pi
    N\to NM}(s_1,t_1)\ ,
\label{eq:AVN}
\end{eqnarray} 
where $p_1^{\prime}$, $p_2^{\prime}$, $p_3^{\prime}$ and
$p_4^{\prime}$ are the 4-momenta of the $\pi$ meson, the initial
nucleon, the final nucleon and the vector meson, respectively. The
$\lambda_i^{\prime}$ are the spin projections of the particles, 
$\vec{\epsilon}^{\,(V)}$ is the polarization vector of the vector meson
and $s_1=(p_1^{\prime} +
p_2^{\prime})^2=(p_3^{\prime}+p_4^{\prime})^2$, $t_1=(p_1^{\prime} -
p_4^{\prime})^2=(p_2^{\prime}-p_3^{\prime})^2$.

The invariant amplitude is normalized to the total cross section as
follows
\begin{equation} 
  | A_{\pi^0 N \to M N}(s_1,t)|^2 = | A_{\pi^0 N \to M N}(s_1,u)|^2 =
  {\frac8 3} \pi s_1
  \frac{p_{\pi}^{\mathrm{cm}}}{p_M^{\mathrm{cm}}}\sigma_ {\pi^- p \to
    M n}
\end{equation}
where $s_1$ is the invariant mass squared of the $Mn$ system. 

It was shown in Ref.\cite{Grishina} that apart from the $\pi$-exchange
contributions heavier vector-meson exchanges --- especially of $\rho$
mesons --- are important for the case of the reactions $p n \to d \eta
$ and $p n \to d \eta^{\prime}$.  In our case the amplitudes for the
vector-meson exchanges can be written in the form
\begin{eqnarray}
  A^V_{pn\to dM}(s,t) &&= \frac{G_V}{2\, m}\varphi^T_{\lambda_2}
  (\vec{p}_2) (-i\, \sigma_2) \cdot A_{V^0 N\to
    MN}(s_1,t) \ \times \nonumber\\
  &&\left\{ \vec{M}_1^V(\vec{p}_1)\cdot\vec{\epsilon}_M^{\ast}
    \vec{\sigma} \cdot \vec{\epsilon}_d^{\ast} +
    \vec{M}_2^V(\vec{p}_1)\cdot\vec{\epsilon}_d^{\ast} \vec{\sigma}
    \cdot \vec{\epsilon}_M^{\ast} \ - \right.  \nonumber\\
  &&\left. \mbox{ }\vec{\sigma}\cdot \vec{M}_2^V(\vec{p}_1)\,
    \vec{\epsilon}_d^{\ast} \cdot \vec{\epsilon}_M^{\ast} + i\left[
      \vec{M}_2^V(\vec{p}_1)\times \vec{\epsilon}_d^{\ast}
    \right]\cdot \vec{\epsilon}_M^{\ast}
  \right\} \varphi_{\lambda_1}(\vec{p}_1) \label{AVt} \\
  A^V_{pn\to dM}(s,u) &&= \frac{G_V}{2\, m}\varphi^T_{\lambda_1}
  (\vec{p}_1) (-i\, \sigma_2) \cdot A_{V^0N\to
    MN}(s_1,u) \ \times \nonumber\\
  &&\left\{ \vec{M}_1^V(-\vec{p}_1)\cdot\vec{\epsilon}_M^{\ast}
    \vec{\sigma} \cdot \vec{\epsilon}_d^{\ast} +
    \vec{M}_2^V(-\vec{p}_1)\cdot\vec{\epsilon}_d^{\ast} \vec{\sigma}
    \cdot
    \vec{\epsilon}_M^{\ast} \ - \right. \nonumber\\
  &&\left. \mbox{ } \vec{\sigma}\cdot \vec{M}_2^V(-\vec{p}_1)\,
    \vec{\epsilon}_d^{\ast} \cdot \vec{\epsilon}_M^{\ast} + i\left[
      \vec{M}_2^V(-\vec{p}_1)\times \vec{\epsilon}_d^{\ast}
    \right]\cdot \vec{\epsilon}_M^{\ast} \right\}
  \varphi_{\lambda_1}(\vec{p}_1) \ , \label{AVu}
\end{eqnarray}
where
\begin{eqnarray}
\vec{M}_1^V(\vec{p}_1)& =& \sqrt{2\, m} 
  \int [ (\vec{k}-\vec{p}_1) +
  \frac{\vec{k}^2-\vec{p}^2_1}{m^2_V} (\vec{k}+\vec{p}_1)]  
\,  \Phi_V(\vec{k},\vec{p}_1)
  \Psi_d^{\ast}(\vec{k}) \frac{\mathrm{d}^3k}{(2\pi)^{3/2}} 
\end{eqnarray}
and
\begin{eqnarray}
  \vec{M}_2^V(\vec{p}_1) &=& \sqrt{2\, m} \int
  (1+\kappa_V)(\vec{k}+\vec{p}_1)\, \Phi_V(\vec{k},\vec{p}_1)
  \Psi_d^{\ast}(\vec{k}) \frac{\mathrm{d}^3k}{(2\pi)^{3/2}} \ .
\end{eqnarray}
The function $\Phi_V(\vec{k},\vec{p}_1)$ describes the product of the
$V$-meson propagator $(q^2 - M_V^2)^{-1}$ and the form factor at the
$VNN$ vertex $F_V(q^2)$. It is defined by Eq.~(\ref{ff}) where
$m_{\pi}^2$ should be substituted by $m_V^2$. $G_V$ and $\kappa_V G_V$
are the vector and tensor coupling constants respectively.

The general spin structure of the $V N \to MN$ amplitude near
threshold has the following form
\begin{eqnarray}
\lefteqn{
  \langle p_3^{\prime}\,\lambda^{\prime}_{3}; p_4^{\prime}
  \lambda_4^{\prime} | \hat{T}_{V N\to NM} | p_1^{\prime}
  \lambda^{\prime}_{1} ;
    p_2^{\prime}\lambda_2^{\prime}\rangle =}  \nonumber \\
    &\varphi_{\lambda^{\prime}_{4}}^{\ast}(\vec{p}_4^{\,\prime})\ 
    &\left( \vec{\epsilon}_{\lambda^{\prime}_{3}}^{\ast \,(M)} \cdot
      \vec{\epsilon}_{\lambda^{\prime}_{1}}^{\,(V)} \ 
      A_{V N\to NM}(s_1,t_1)\ + \right.\nonumber \\
    &&\left.i \left[ \vec{\epsilon}_{\lambda^{\prime}_{3}}^{\ast \,(M)}
        \times \vec{\epsilon}_{\lambda^{\prime}_{1}}^{ \,(V)}\right]
      \cdot \vec{\sigma} \ B_{V N\to NM}(s_1,t_1)\right)
    \varphi_{\lambda^{\prime}_{2}}^{\ast}(\vec{p}_2^{\,\prime})\ ,
\label{eq:AVM}
\end{eqnarray} 

where the notations are similar to the ones in Eq.(\ref{eq:AVN}).  Two
invariant amplitudes $A_{V N\to NM}(s_1,t_1)$ and $B_{V N\to
  NM}(s_1,t_1)$ are necessary to describe two possible transitions
$\left(\frac{1}{2}\right)^- \to \left(\frac{1}{2}\right)^-$ and
$\left(\frac{3}{2}\right)^- \to \left(\frac{3}{2}\right)^-$.  It is
known from the data on Compton scattering (see, e.g., \cite{Erbe})
that the spin-flip amplitude $B_{\gamma N\to \gamma N}(s_1,t_1)$ is
small as compared with the non spin-flip amplitude $A_{\gamma N\to
  \gamma N}(s_1,t_1)$ except in the $\Delta$-resonance region (see,
e.g., \cite{Erbe}).  Following the arguments of the Vector-Dominance
Model (VDM) we assume that this amplitude is also small in our case
and take into account only the first non spin-flip term in
Eq.(\ref{eq:AVM}).

Note that the amplitudes $A^{\pi}$ and $A^{\rho}$ correspond to the
exchange of neutral $\pi$ and $\rho$ mesons only (see the left
diagrams in Fig.~\ref{fig:tsm}). To take into account also the charged
$\pi$ and $\rho$ exchanges we have to multiply amplitude (\ref{Atu})
by a factor 3. Of course in the case of $\omega$ exchange such a
factor is not necessary.

Therefore, the differential cross section of reaction (\ref{eq:Vd})
can be written as
\begin{eqnarray}
  \lefteqn{ \frac{\d\sigma_{pn\to dM}}{\d t} = }
  \nonumber \\
  && \frac{1}{64\,\pi s}\ \frac{1}{(p_{1}^{\mathrm{cm}})^2}\ F(I)\ 
  \overline{|A_{pn \to dM}(s,t) + A_{pn \to dM}(s,u) |^2} \ .
\label{eq:sigmaVd}
\end{eqnarray}
where the isospin factor $F(I)$ is equal to 9 for isovector exchanges
($\rho$ and $\pi$) and 1 for isoscalar exchange ($\omega$).

\section{Choice of parameters and results of calculations}
\label{sec:results}
We assume the form factors $F_{\pi}(q^2)$ and $F_{V}(q^2)$ to be of
monopole type.  Recent QCD lattice calculations \cite{QCD} suggest
that the cut-off in the pion form factor should be quite soft
$\Lambda_\pi \simeq 0.8 $ GeV/c (see also Refs.~\cite{Coon,Speth2}).
Of course such a soft pion form factor suppresses pion exchange and
contributions of heavier meson exchanges become more important. This
for example was demonstrated in Ref.~\cite{Grishina} where it was
found that the $\rho$-exchange contribution in the reactions $p n \to
d \eta$ and $p n \to d \eta^{\prime}$ is significant.  Here also
$\Lambda_\pi = 0.8 $ GeV/c is used.

The coupling constants and vertex form factors for $\rho$
and $\omega$ mesons are taken from the full Bonn $NN$
potential \cite{Holinde}: $G_{\rho}^2/4\pi = 0.84$, $\kappa_\rho =
6.1$, $G_{\omega}^2/4\pi = 20$, $\kappa_{\omega} = 0$ and
$\Lambda_\rho$ = 1.4 GeV/c, $\Lambda_\omega$ = 1.5 GeV/c.

For the deuteron wave function we take the parameterization from
Ref.\cite{Lacomb} and neglect the $D$-wave part. As it was
demonstrated in Ref.\cite{kondrat2} for the case of the reaction $\bar
p d \to M n$ (where the same structure integrals (\ref{M}) for $\pi$,
$\rho$ and $\omega$ exchanges occur) the $D$-wave term of the deuteron
wave function gives a negligibly small contribution compared
to the $S$-wave term.

To define the amplitudes $\pi N \to M N$ we use the following values
of the $S$-wave cross sections (taken from
Ref.\cite{Binnie}):\\
$\sigma_{\pi^- p \to \omega n} = (8.3 \pm 0.07) p^M_{\mathrm{cm}}\ 
\mu$b and $\sigma_{\pi^- p \to \phi n} = (0.29 \pm 0.06)
p^M_{\mathrm{cm}}\ \mu$b$\\$ ($p^M_{\mathrm{cm}}$ in MeV/c).  The
experimental data show that the angular distribution in the reaction
$\pi^- p \to n\omega$ is isotropic and the $S$-wave is dominant at
least up to $k_V^{\mathrm{cm}}(s_1) = 260$ MeV/c (see the comment on
p.2805 in \cite{Binnie}).  We ignore an apparent suppression of the
$S$-wave amplitude very close to threshold ($k_V^{\mathrm{cm}}(s_1)
\leq 80 - 100$ MeV/c), reported in Ref.\cite{Binnie}, because
according to Ref.\cite{Hanhart} this effect has a kinematical origin.

The contributions from the $\rho$ and $\omega$ exchanges are
calculated using the vector-dominance model (VDM) prediction for the
amplitude $\rho N\to \omega(\phi) N$ and assuming that for
non-diagonal cases $A_{\omega N\to M N} \approx A_{\rho^0 N\to M N}$.
We derive the $S$-wave $\gamma N\to \omega N$ amplitude from the
ABBHHM data at $E_{\gamma} = 1.3$~GeV (see Ref.\cite{Erbe}) using a
value of the cross section of the reaction $\gamma p\to \omega p$
equal to 5.6 -- 7.8 $\mu$b. This would give the $\rho p\to \omega p$
cross section of about 2.7 $\pm$ 0.5 mb at low energies. The ratio of
the $\gamma p\to \phi p$ and $\gamma p\to \omega p$ amplitudes squared
was found from the data at $s$ = 5 -- 6 GeV$^2$ to be 0.06 -- 0.07.
Then we assumed that it is the same for the the case of the reactions
$\rho p\to \phi p$ and $\rho p\to \omega p$.  For the elastic $\omega
N$ scattering cross section at low energies we took the value 15 mb
which was evaluated in Ref.\cite{Lykasov} within the sigma-exchange
model and is in agreement with previous estimations made using the
Quark Model.

Since the relative phases of the different contributions are not known
we calculate the cross section of the reaction $pn\to dM$ as the
incoherent sum
\begin{equation} 
  \sigma_{pn\to dM} = N [\sigma^{(\pi)}+\sigma^{(\rho)}+
  \sigma^{(\omega)}] \ .
\end{equation} 

\begin{figure}[htb]
  \begin{center}
  \leavevmode
  \psfig{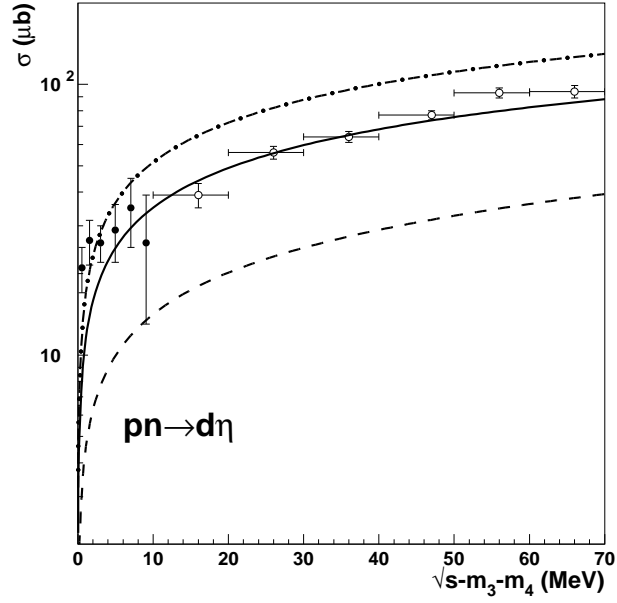}
  \caption{Cross section of the reaction $pn \to d \eta$ 
    as a function of the c.m.\ excess energy (taken from
    Ref.\cite{Grishina}).  The dashed curve shows the $\pi$-exchange
    contribution whereas the dash-dotted curve is the sum of $\pi$,
    $\rho$, and $\omega$ exchanges.  The solid curve includes all
    contributions ($\pi$, $\rho$, $\omega$) multiplied with a
    normalization factor $N =0.68$ in order to take into account
    effects from the initial state interaction (see text).  The data
    points for are taken from Refs.~\protect\cite{Cal1} (open circles)
    and \protect\cite{Cal2} (filled circles).}
   \label{fig:eta}
  \end{center}
\end{figure}

In Fig.\ref{fig:eta} taken from Ref.\cite{Grishina} we show how
the TSM (with the same coupling constants and cut-off parameters for
$\pi$, $\rho$ and $\omega$ exchanges and the $S$-wave amplitudes
$Vp \to \eta p$ and $ Vp \to \eta^{\prime}p$ estimated using VDM from
the photo-production data) describes the experimental data on the
reaction $p n \to d \eta$.  The cross section of the reaction $pn \to d
\eta$ is presented as a function of the c.m.\ excess energy $Q$. The
dashed curve shows the $\pi$-exchange contribution alone whereas the
dash-dotted curve describes the sum of $\pi$, $\rho$, and $\omega$
exchanges. The solid curve includes all contributions ($\pi$, $\rho$,
$\omega$) multiplied with a normalization factor $N =0.68$ in order to
take into account effects from the initial state interaction (ISI).
The data points for are taken from Refs.\cite{Cal1} (open
circles) and \cite{Cal2} (filled circles).  The reduction
factor appeared to be not very different from the prediction of the
ISI effect within a simple model which assumes the dominant
contribution from the on-shell rescattering \cite{Nakayama} and gives
$\lambda_{\mathrm{ISI}} \simeq 0.5$.

As we see from Fig.\ref{fig:eta} pion exchange calculated with the soft
cut-off parameter cannot describe the $\eta$-production data and the
contribution from heavier meson exchanges (and especially of $\rho$
\cite{Grishina}) is quite important.

In Figs.\ref{fig:om} and \ref{fig:phi} we present the predictions of
the TSM for the cross sections of the $\omega$ and $\phi$ production.
The contribution of pion exchange is shown by the dashed curves.  The
lower and upper curves show the minimal and maximal values of the
$\pi$-exchange contribution demonstrate which follow from the
experimental errors of the elementary cross sections.  The dash-dotted
curves describe the sum of $\pi$-, $\rho$- and $\omega$-exchange
contributions.  The solid curves represent the results including all
contributions ($\pi$, $\rho$, $\omega$) multiplied with the same
normalization factor $N =0.68$ as in the case of $\eta$- production in
order to take into account effects from ISI. It is clearly seen that
similar to the case of $\eta$ production the $\rho$-exchange
contribution to the cross sections of the reactions is very
significant.  The relative contribution of $\pi$ exchange is about 20
\% in the case of $\omega$ production and is almost 2 times less in
the case of $\phi$ production. The $\omega$ exchange is more
important in the case of $\omega$ production where it gives about
20\%; in the case of $\phi$ productions its relative contribution is
about 5\%.

The cross sections of the reactions $p n \to \omega d$ and $p n \to
\phi d$ can be parameterized as follows
\begin{equation} 
  \label{pndM}
  \sigma_{pn\to d M} \approx D_{M}\sqrt{Q}\:,
\end{equation} 
where $D_{\omega} =( 2.7\pm 0.3)$~$ \mu$b/MeV$^{1/2}$ and $D_{\phi} =
(0.09 \pm 0.02)$~$\mu$b/MeV$^{1/2}$ . At very low Q which are of the
order of the resonance width each cross section might be a little
larger because of the finite widths of the $\omega$ and
$\phi$~\cite{SPES3}.

In Fig.\ref{fig:om} we show also experimental data on the
near-threshold production of $\omega$ mesons in the $pp \to pp \omega$
reaction \cite{SPES3}.  Near threshold the predicted cross section of
$\omega$ production with the deuteron in the final state is much
higher than that of the reaction the $pp \to pp \omega$.  This is very
similar to the case of $\eta$ production (see , e.g.,  
\cite{Cal1,Cal2}) and is related to isospin and phase-space factors
(see, e.g., \cite{Wilkin98}).

Let us discuss the relation between $\sigma(pp\to pp \omega)$ and
$\sigma(pn\to d \omega)$ near threshold in more detail.  F\"aldt and
Wilkin \cite{Faldt} proposed the following parameterization of the
cross section of the reaction $pp \to pp M$ near the threshold
\begin{equation} 
  \label{FW} 
  \sigma_{pp\to pp M} = C_{M}\left(\frac{Q}{\epsilon}\right)^{\!{2}}
  \left(1+\sqrt{1+Q/\epsilon}\,\right)^{\!-2}\:.
\end{equation} 
This formula takes into account the strong final state interaction of
two protons including also Coulomb distortion with $\epsilon\approx
0.45$~MeV. For $\eta$ and $\omega$ production we have $C_{\eta} = (110
\pm 20)$~nb and $C_{\omega} = (37 \pm 8)$~nb ~\cite{SPES3}. At $Q$=15
MeV we have $\sigma(pp\to pp \eta) \approx 2.6\ \mu$b ($\sigma(pp\to
pp \omega) \approx 1\ \mu$b) which is 15(10) times less than the
cross section of the reaction $pn\to d \eta$ ($pn\to d \omega$).  Note
that in line with suggestions by Wilkin (see, e.g., \cite{Wilkin98}) the
ratios $\sigma(pn\to d \eta)/\sigma(pp\to pp \eta)$ and
$\sigma(pn\to d \omega)/ \sigma(pp\to pp \omega)$ are, in fact, not
very different.

The reaction $ p p \to p p \omega$ near the threshold was also
analyzed within the framework of the meson-exchange model in
Ref.\cite{Speth}.  Adjusting the cut-off parameter of the form factor
to the low energy data the authors of Ref.\cite{Speth} calculated the
cross section of the reaction $ p p \to p p \omega$ for proton
incident energies up to 2.2 GeV. This model predicts a cross section
of about 15--20~$\mu$b at $Q \approx 100 $ MeV which is still not very
different from parameterization (\ref{FW}).  If parameterizations
(\ref{pndM}) and (\ref{FW}) would be valid up to $Q= 1$ GeV then the
cross section of the reaction $\sigma(pp\to pp \omega)$ would reach
the same value as the cross section of the reaction $pn\to d \omega$
only at 900 MeV.  Of course those formulas can not be valid up to such
large values of Q. Estimations within the framework of the Quark-Gluon
String Model shows that the cross section of the reaction $pn\to d
\omega$ can reach maximum of about 30--50~$\mu$b at $Q$ = 100--200 MeV
and then will start to fall (see \cite{proposal}). According to the
parameterization of Ref.\cite{Sibirtsev} the cross section of the
reaction $\sigma(pp\to pp \omega)$ reaches the value of $30\ \mu$b at
$Q\approx 200$ MeV.  Therefore we can expect that in a rather broad
interval of $Q$ (at least up to about 100--150 MeV) the cross section
of the reaction $pn\to d \omega$ will be larger than the cross section
of the reaction $\sigma(pp\to pp \omega)$. This gives quite a good
chance that the reaction $pn\to d \omega$ can be detected using
missing mass method at COSY by measuring the forward deuteron and
spectator proton in the reaction $p d \to d \omega p_{\mathrm{sp}}$.

For the case of $\phi$ production we also expect that near
threshold the cross section of the reaction $pn\to d \phi$ will be
larger than the cross section of the reaction $pp \to pp \phi$. The
latter was estimated using DISTO data in Ref.\cite{Wilkin98} and found
to be equal to 0.28$\pm$0.14 $\mu$b at~$Q=82$~MeV.  Though there are
uncertainties in extrapolating the prediction of the TSM
(Eq.(\ref{pndM})) to such large $Q$ we would have $\sigma(pn\to d
\phi) \approx 0.6 -1\ \mu$b at this $Q$.

Let us discuss now the $\phi/\omega$ ratio. TSM predicts the following
value
\begin{equation} 
  \label{ratio}
  R_{pn \to d M} =D_{\phi}/D_{\omega}=(30 \pm 7)\times 10^{-3}.
\end{equation} 
This is lower than the corresponding ratio in $pp$ collisions
\cite{SPES3}
\begin{equation}
  \label{ratio2}
  R_{pp \to ppM} =C_{\phi}/C_{\omega}=(49 \pm 26)\times 10^{-3}.
\end{equation}
and in the reaction $pd \to ^{3}\!\!He\,M$ (see Eq.(3)).  It is
closer to the ratio of the $\phi$ to $\omega$ yields in $\pi^- p$
collisions (see, e.g., the discussion in Ref.\cite{Wilkin98})
\begin{equation}
  \label{ratio3}
  R_{\pi^-p  \to nM} =(37 \pm 8)\times 10^{-3}.
\end{equation}
Another estimate of $R$ can be found if we assume the line-reverse
invariance of the amplitudes, which correspond to the diagrams
presented in Fig.\ref{fig:tsm}.  In this case we have 

\begin{eqnarray}
  \overline{
    |T^{\mathrm{LRI}}_{pn \rightarrow dM}(s,t)|^2} & = &
    \overline{
      |A^{\mathrm{LRI}}_{pn \rightarrow dM}(s,t) +
      A^{\mathrm{LRI}}_{pn \rightarrow dM}(s,u)|^2} 
    \nonumber \\
    & = &
    \overline{
      |A_{\bar{p}d \rightarrow nM}(s,t) +
      A_{\bar{p}d \rightarrow nM}(s,u)|^2}
\end{eqnarray}    
  
and can define the ratio
\begin{equation}
  R_{\mathrm{LRI}}=|T^{\mathrm{LRI}}_{pn \to d \phi}|^2/
  |T^{\mathrm{LRI}}_{pn \to d \omega}|^2=
  |T_{\bar{p}d \to n \phi}|^2/
  |T_{\bar{p}d \to n \omega}|^2.
\end{equation}

Adopting the result of the OBELIX collaboration $ Y(\bar{p}d \to
n\phi)/ Y(\bar{p}d \to n\omega) = (230 \pm 60)\times 10^{-3}$ we get
\begin{eqnarray}
  R_{\mathrm{LRI}} & = &|T_{\bar{p}d\to n \phi}|^2/|T{\bar{p}d \to n
    \omega}|^2
  \nonumber \\
  & \approx & (p_{\mathrm{cm}}^{\omega}/p_{\mathrm{cm}}^{\phi})
  (Y(\bar{p}d \to n \phi)/ Y(\bar{p}d\to n\omega)) \simeq (250 \pm
  60)\times 10^{-3},
\end{eqnarray}
which is larger by an order of magnitude than the prediction of the
TSM given by Eq.(\ref{ratio}).  If experimental studies will find an
essential excess of $R(\phi /\omega)$ over the value predicted by the
two-step model it might be interpreted as a possible contribution of
the intrinsic $s\bar{s}$ component in the nucleon wave function.

\section{Conclusions}
\label{sec:conc}
Using the two-step model which is described by triangle graphs with
$\pi$-, $\rho$- and $\omega$-meson exchanges we calculated the cross
sections of the reactions $pn\to dM$, where $M=\omega$ or $\phi$,
close to threshold. The predicted cross section of the reaction $pn\to
d\omega$ is found to be significantly larger than the cross section of
the reaction $pp\to pp\omega$. The same is expected to be the case for
$\phi$ production.  We find a $\phi/\omega$ ratio of $R_{pn \to
  dM}=(30 \pm 7)\times 10^{-3}$.  The measurement of the $\phi$ and
$\omega$ yields in the reaction $pn\to dM $ at the same energy release
$Q$ will be useful for a better understanding of the mechanism of the
OZI-rule violation.

\begin{ack}
  We are grateful to W.\ Cassing, Ye.S.\ Golubeva, M.G.\ Sapozhnikov
  and C.\ Wilkin for useful discussions.
\end{ack}

\newpage

\normalsize

\clearpage

\clearpage

\begin{figure}[htb]
  \begin{center}
  \leavevmode
  \psfig{file=fig3.epsi,width=8.cm}
  \caption{Cross section of the reaction $pn \to d \omega$
    as a function of the c.m.\ excess energy.  The dashed curves show
    the $\pi$-exchange contribution alone whereas the dash-dotted
    curves are the sums of $\pi$, $\rho$, and $\omega$ exchanges.  The
    solid curves include all contributions ($\pi$, $\rho$, $\omega$)
    multiplied with a normalization factor $N=0.68$ in order to take
    into account effects from the initial state interaction (see
    text).  The upper and lower dashed, solid and dash-dotted curves
    are the results obtained using the maximal and minimal values of
    the elementary $\pi N \to \omega N$ and $V N \to \omega N$
    $S$-wave amplitudes (see text).  The data points are the data on
    the reaction $pp \to pp \omega$ from Ref.\protect\cite{SPES3}.}
   \label{fig:om}
  \end{center}
\end{figure}

\clearpage

\begin{figure}[htb] 
  \begin{center}
    \leavevmode
    \psfig{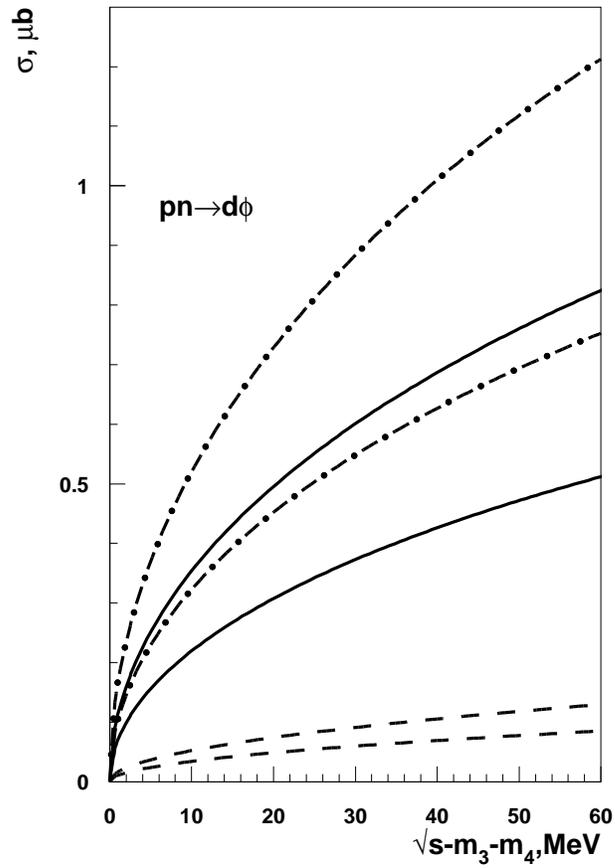}
    \caption{Cross section of the reaction $pn \to d \phi$
      as a function of the c.m.\ excess energy. The meaning of the
      curves is the same as in Fig.\ref{fig:om}.}
    \label{fig:phi} 
  \end{center}
\end{figure} 

\end{document}